**Linton Stereo Illusion**

Paul Linton[1,2,3,*]

[1] NOMIS Foundation Fellow, Italian Academy for Advanced Studies, Columbia University
[2] Presidential Scholar in Society and Neuroscience, Center for Science and Society, Columbia University
[3] Visual Inference Lab, Zuckerman Mind Brain Behavior Institute, Columbia University

We present a new illusion that challenges our understanding of stereo vision. The illusion consists of a larger circle at 50cm, and smaller circle in front of it at 40cm, with constant angular sizes throughout. We move the larger circle forward by 10cm (to 40cm) and then back again (to 50cm). The question is, what distance should we move the smaller circle forward and back to maintain a constant perceived separation in depth between the circles? Constant physical distance (10cm) or constant retinal disparity (6.7cm)? Observers choose constant disparity. We therefore argue the 'Linton Stereo Illusion' appears to suggest that perceived stereo depth reflects retinal disparities rather than 3D geometry.

When do two circles appear to move rigidly in depth together? In the 'Linton Stereo Illusion' we test this question by moving a back circle (starting at 50cm) and a front circle (starting at 40cm) forwards and backwards, whilst we keep their angular size constant throughout. We move the larger circle forward by 10cm (to 40cm) and then back again (to 50cm). We move the smaller circle forward to one of two different distances – 30cm in a 'constant physical separation' condition, and 33.3cm in a 'constant retinal disparity' condition – and then back again. The question is, in which of the two conditions do the two circles appear to move rigidly in depth together?

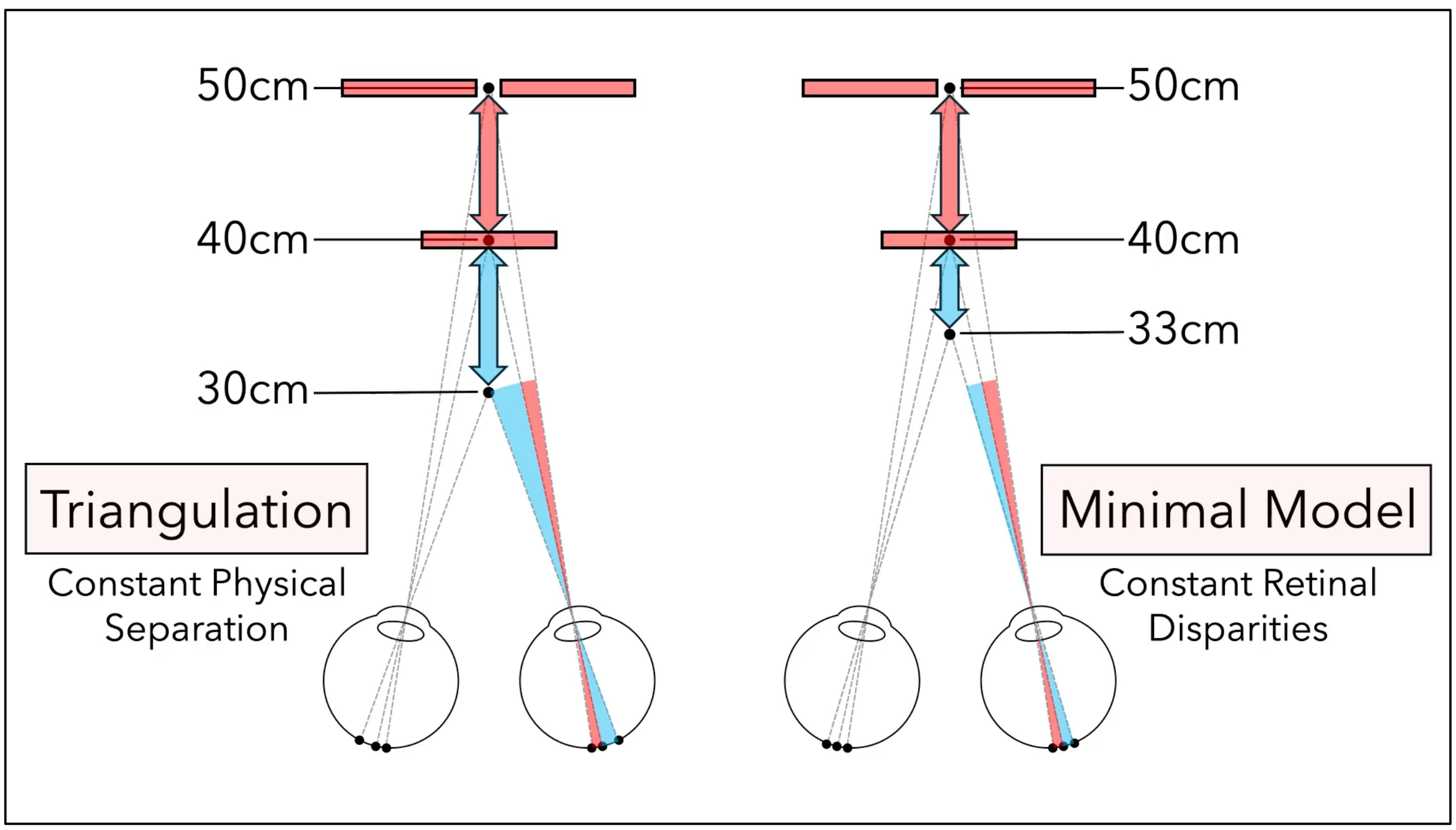

[Insert Movie 1 here: https://youtu.be/wLFn6pfoTg8]

Movie 1. The two conditions of the 'Linton Stereo Illusion'.

* Contact email: paul.linton@columbia.edu

The two conditions reflect two very different ways of thinking about stereo vision:

First, 'Triangulation' models of stereo vision – that date back to (Kepler, 1604) and (Descartes, 1637) – suggest that stereo vision recovers the physical geometry. In which case, we would expect to see the circles moving rigidly in depth together when we keep the physical distance between them constant.

By contrast, a second approach (Linton, 2021b, 2023), argues that perceived stereo depth simply reflects the disparities on the retina. In which case, we would expect to see the circles moving rigidly in depth together when we keep the retinal disparities between them constant, by compressing the physical distance between them as they move forwards in depth (as illustrated in Movie 1).

The vast majority of observers pick the 'constant retinal disparity' condition as the one that moves rigidly in depth. By contrast, in the 'constant physical separation' condition, the near circle appears to move more than the far circle, producing a non-rigid 'concertina' effect (Movie 2).

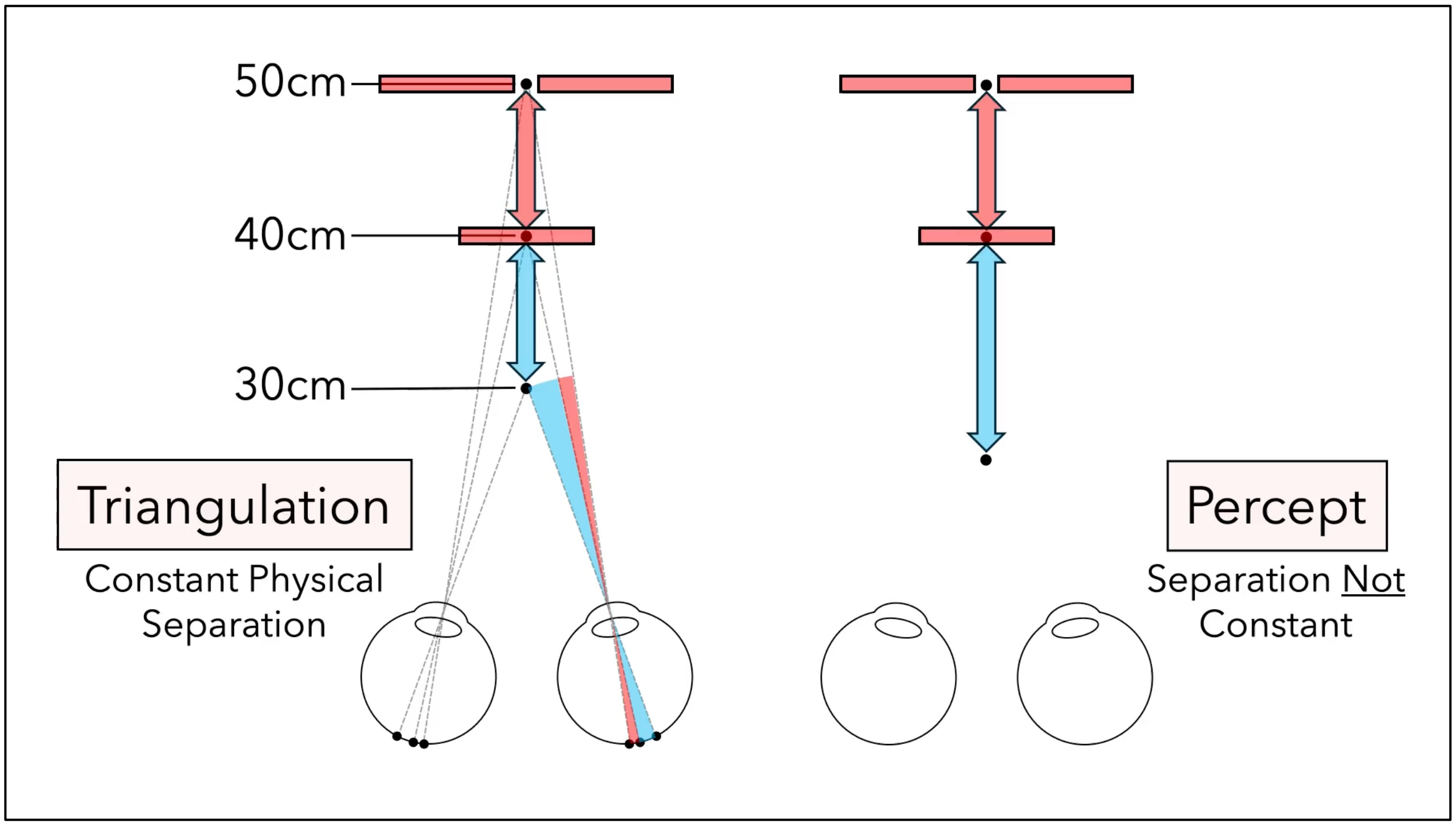

[Insert Movie 2 here: https://youtu.be/qgbAnxt1Kyc]

Movie 2. Perceptual experience (right) in the 'constant physical separation' condition.

Whilst in the 'constant retinal disparities' condition the circles do appear to move pretty much rigidly in depth (Movie 3).

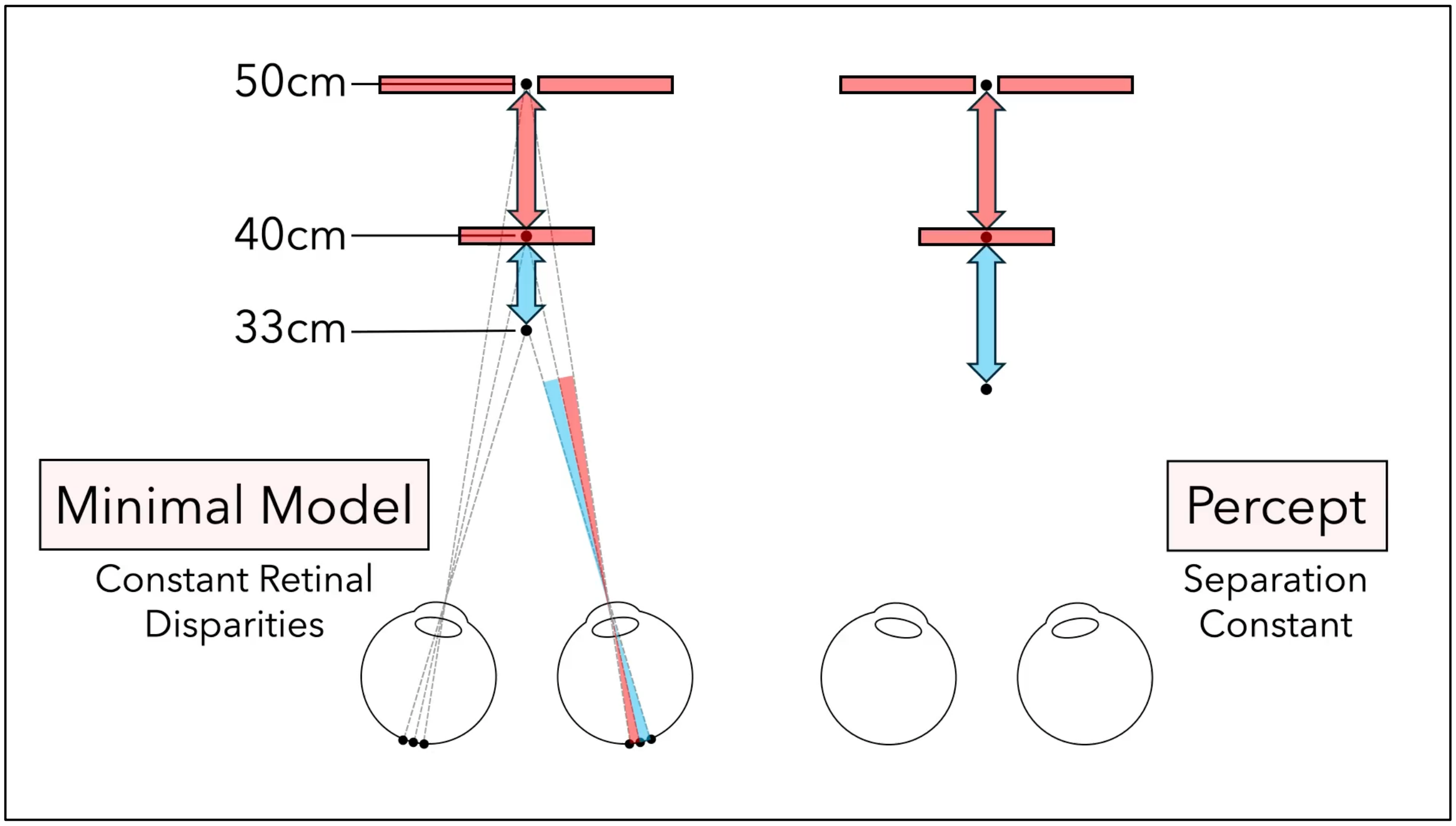

[Insert Movie 3 here: https://youtu.be/P1Aw-Jo5cMk]

Movie 3. Perceptual experience (right) in the 'constant retinal disparities' condition.

The 'Linton Stereo Illusion' was presented at the Applied Vision Association (AVA) Spring 2024 Meeting (Linton & Kriegeskorte, 2024a) and European Conference on Visual Perception (ECVP) 2024 (Linton & Kriegeskorte, 2024b), as well as at the Vision Sciences Society (VSS) 2024 Demo Night and the ECVP 2024 Illusion Night. Roughly ≈210 visitors experienced the 'Linton Stereo Illusion' across the two conferences (≈160 at VSS, and ≈150 at ECVP), with ≈90% of observers picking the 'constant retinal disparities' condition as the rigid one[1]. The 'Linton Stereo Illusion' has also been widely discussed online, receiving 16,500 views on Twitter/X[2].

[1] ≈90% is higher than I expected, given roughly 1 in 8 people are thought to have some form of stereo vision deficits. Although when people mentioned they didn't see any depth, I didn't ask them to decide one way or another.

[2] https://x.com/LintonVision/status/1788572440034947469 (May 2024, 16,500 views as of June 2025).

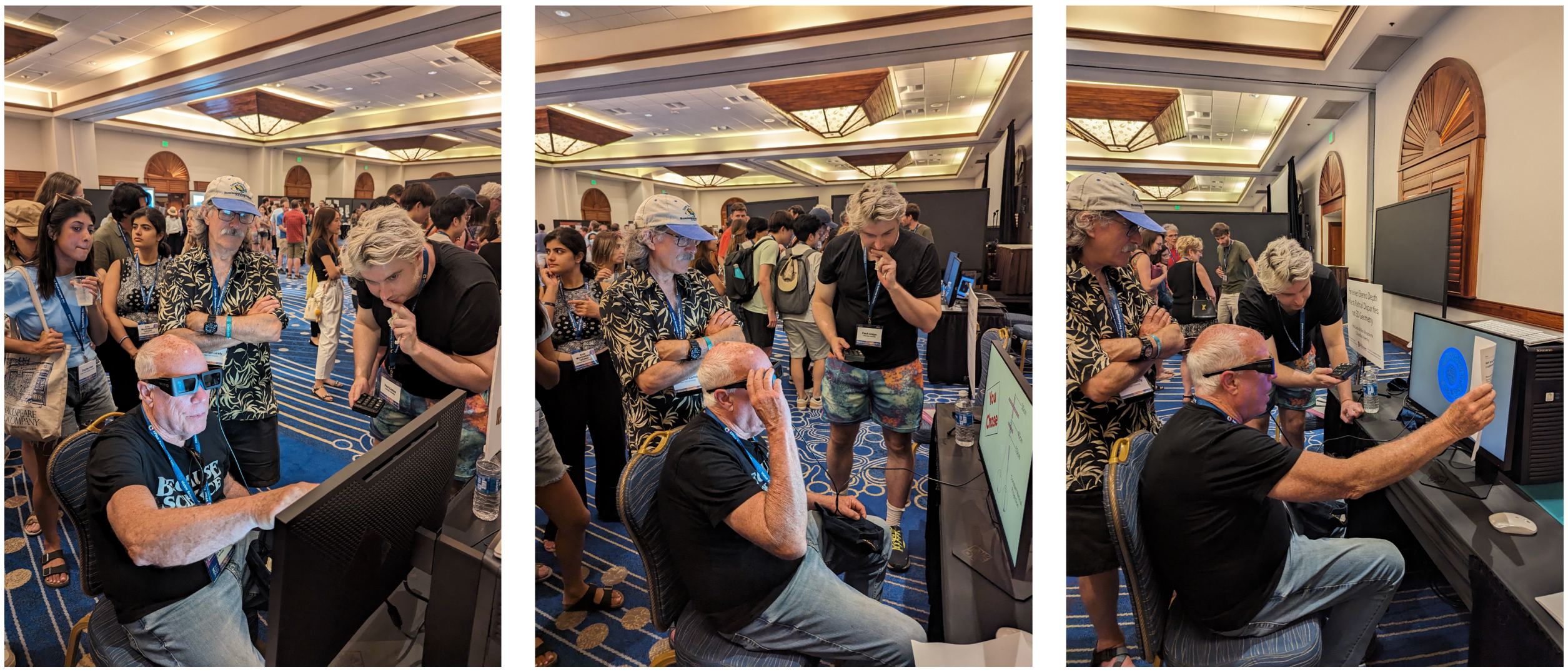

Figure 1. 'Linton Stereo Illusion' at VSS Demo Night 2024. Here being viewed by Marty Banks (University of Berkeley) and Mike Landy (New York University). (Photos by Benjamin Peters).

**Demos**

You can try the 'Linton Stereo Illusion' for yourself in Demo 1 using red-blue stereo glasses. On the left is the 'constant physical separation' condition. On the right is the 'constant retinal disparity' condition. In which of the two conditions do the circles appear to move rigidly?

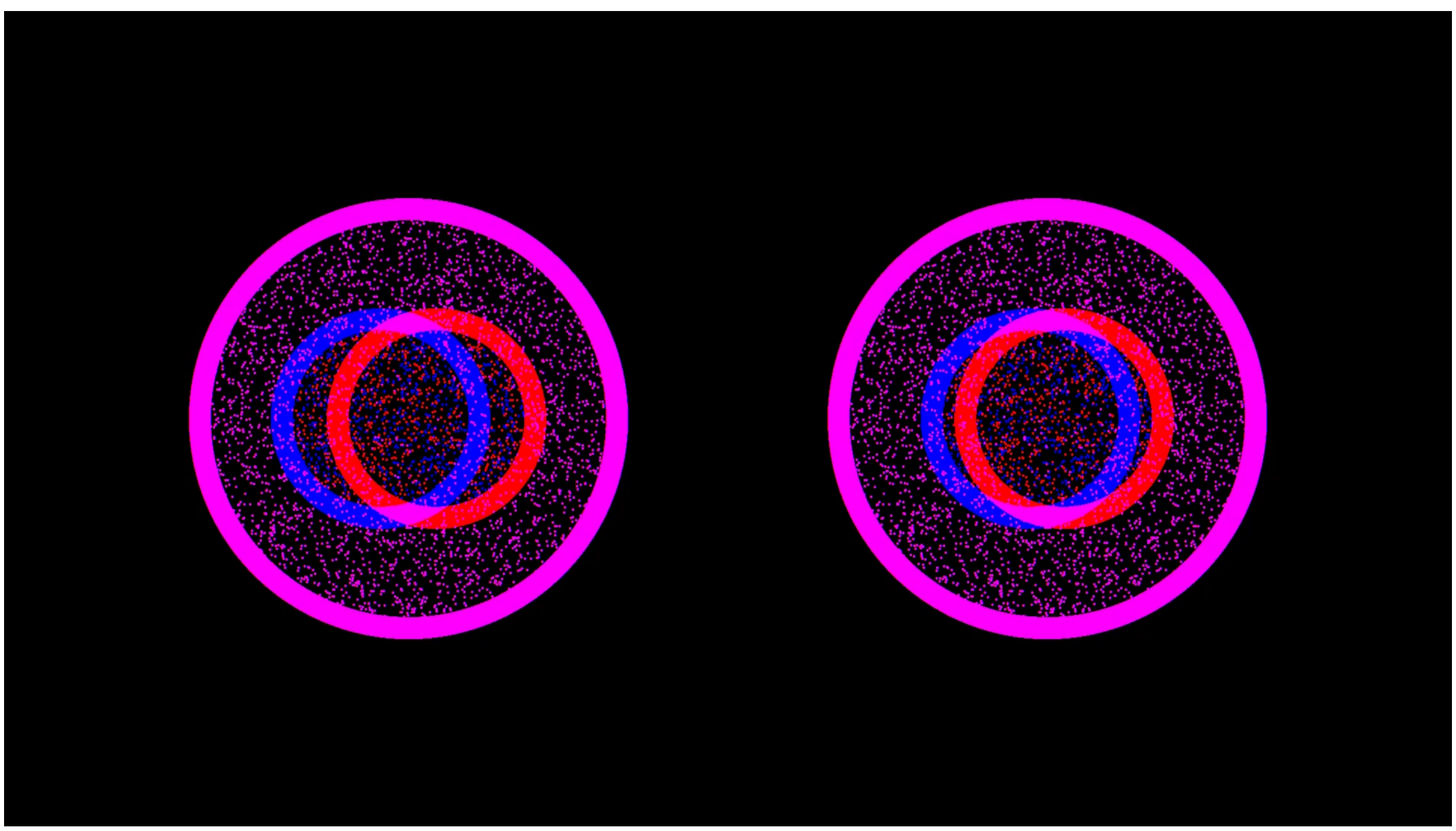

[Insert Movie 4 here: https://youtu.be/beefk6GkBWo]

Movie 4. 'Linton Stereo Illusion' Demo 1: 'Constant physical separation' condition (left) and 'constant retinal disparities' condition (right). Stimuli specified for a 27-inch display at a 40cm viewing distance with 6.4cm interpupillary distance. Demos available at: https://Github.com/StereoIllusion/

Whilst Demo 1 makes it clear how the stimuli work, some readers might see the effect more clearly in Demo 2, which takes the same stimulus, but optimizes it for red-blue glasses.

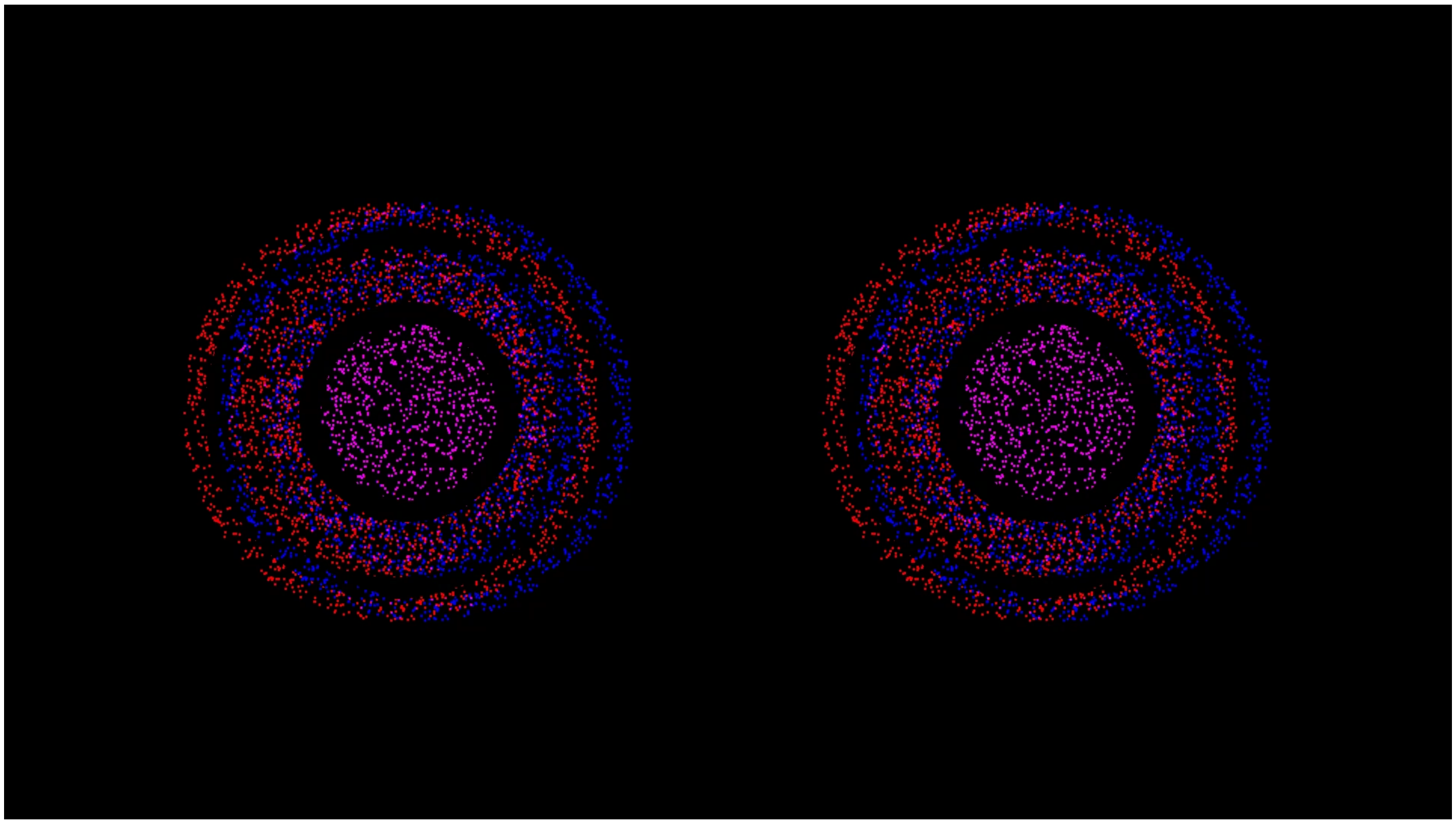

[Insert Movie 5 here: https://youtu.be/zR3Y1pgnQHI]

Movie 5. 'Linton Stereo Illusion' Demo 2: 'Constant physical separation' condition (left) and 'constant retinal disparities' condition (right). Stimuli specified for a 27-inch display at a 40cm viewing distance with 6.4cm interpupillary distance. Demos available at: https://Github.com/StereoIllusion/

There are also ThreeJS (WebGL) and PsychToolbox (Matlab or Octave)(Kleiner et al., 2007) versions of all the demos in this paper, with easily amendable variables like interpupillary distance (IPD), viewing distance, separation between the circles, and speed of motion. The ThreeJS (WebGL) demos run in a normal web browser and will automatically scale appropriately to the screen size.

Project Webpage: https://StereoIllusion.Github.io

Project GitHub: https://Github.com/StereoIllusion/

The 'Linton Stereo Illusion' has also been presented as one of 'Five Illusions that Challenge Our Understanding of Visual Experience' (Linton, 2025) at the Vision Sciences Society (VSS) 2025 (and VSS Demo Night), European Conference on Visual Perception (ECVP) 2025 (and ECVP Demo Night), Cognitive Computational Neuroscience (CCN) 2025, and Optica Fall Vision Meeting 2025.

Project Webpage: https://FiveIllusions.Github.io/

Project GitHub: https://Github.com/FiveIllusions

**What's the Significance of the 'Linton Stereo Illusion'?**

The experience of the 'Linton Stereo Illusion' is undeniable. The question then turns to its significance. I will consider three potential explanations:

1. First, my own explanation, namely that perceived stereo depth reflects retinal disparities rather than 3D geometry (Linton, 2023, 2024a).

And then two alternative explanations that have been put to me:

2. Second, that the illusion demonstrates a well-known failure of 'depth constancy', caused by the visual system scaling disparities with a faulty estimate of the viewing distance (Johnston, 1991).

3. Third, that the illusion is simply an artefact of keeping the angular size of the stimulus (the size of the stimulus on the retina) constant.

Although I would estimate a significant majority of vision scientists that I discussed the 'Linton Stereo Illusion' with at VSS and ECVP agreed with my interpretation, there were notable exceptions, and I cannot claim there was a clear consensus. Hopefully this article, along with the demos at VSS and ECVP, will spur an open, lively, and public exchange of ideas in print. And I very much look forward to – and indeed encourage – challenges to the explanation that I outline here.

That being said, the 'Linton Stereo Illusion' is a clear prediction of my 'Minimal' (disparity-only) account of stereo vision in (Linton, 2023). As (Popper, 1963) suggests, the predictive power of a theory has to seriously count in its favour. Indeed, the challenge facing alternative explanations is significant. Observers in the 'Linton Stereo Illusion' act as if they have direct access to retinal disparities (I would argue, because they do). So alternative explanations of the illusion have to explain why observers act as if they have direct access to disparities, without conceding that they do.

Our field has benefitted immensely from the phenomenal exploration of illusions, from Wheatstone's original papers (Wheatstone, 1838, 1852), to (Helmholtz, 1866), to more recent phenomena like the 'random-dot stereogram' (Julesz, 1960), the 'hollow-face illusion' (Gregory, 1970), and 'Adelson's chequerboard' (Adelson, 1995). No one would deny they have all had a profound effect on our field, even though they are simply phenomenal reports. And subsequent experimental study has only really served to parameterize the fundamental insights we already get from the illusions.

It is important to have this discussion now as a 'Phenomenal Report' for three reasons. First, the 'Linton Stereo Illusion' is already being widely discussed in the vision science community (at VSS, ECVP, and online[3]), and many vision scientists have asked me to fully describe the illusion and explain my interpretation so they can better understand it. Second, I've been struck that when leading vision scientists have offered alternative explanations, they have been so divergent, and so it's important to explore these alternative explanations as well. Third, this dialogue helps to clarify what actual experiments would be worth doing down the line in order to adjudicate between these competing interpretations. As you will see, each potential explanation rests on a number of assumptions, and it would be good to clarify where they disagree in advance of experimental investigation.

[3] https://x.com/LintonVision/status/1788572440034947469 (May 2024; 16,500 views as of June 2025).

## 1. My Explanation: Perceived Stereo Depth Reflects Retinal Disparities

In the published abstracts of the 'Linton Stereo Illusion' for the Applied Vision Association (AVA) Spring 2024 Meeting (Linton & Kriegeskorte, 2024a) and European Conference on Visual Perception (ECVP) 2024 (Linton & Kriegeskorte, 2024b), I draw four specific conclusions:

*1. "First, perceived stereo depth appears to be best captured by the 'Minimal Model'."*

The 'Linton Stereo Illusion' is intended to adjudicate between traditional 'Triangulation' accounts of stereo vision and my own 'Minimal' (disparity-only) theory of stereo vision (Linton, 2023).

You might argue that the two conditions only test 'full depth constancy' accounts like (Guan & Banks, 2016) vs. 'disparity-only' accounts (Linton, 2023), but leave untouched 'partial depth constancy' accounts like (Johnston, 1991). In the next section, I will argue that this is a mistake. I argue that no plausible 'Triangulation' account predicts the pattern of distortions that we experience in the 'constant physical separation' condition of the 'Linton Stereo Illusion'.

*2. "Second, doubling disparity appears to double perceived depth, suggesting that perceived stereo depth is proportional to disparity."*

If the stimulus in the 'constant retinal disparity' condition is perceived as roughly rigid, then the 33-40cm separation and the 40-50cm separation, have (a) the same disparity, and (b) the same perceived depth. If this is true, then when both separations are added together, the new separation (33-50cm) must have (a) double the disparity, and (b) double the perceived depth.

*3. "Third, changes in vergence appear to have no effect on perceived depth."*

The circles appear to move rigidly in depth in the 'constant retinal disparity' condition when vergence tracks the stimulus. It therefore becomes difficult to argue, as (Banks, 2000) does, that:

> "We have shown … that changes in the eyes' vergence can cause a compelling change in perceived shape even when the retinal disparities are completely constant."

Instead, following my work on vergence as a distance (Linton, 2020) and size cue (Linton, 2021a), a new position is emerging in the literature. (Rogers, 2023) cites my work, and concludes that it:

> "…it raises the question of whether there is any role for the vergence angle of the eyes in the perception of spatial layout in natural scenes (Linton, 2021a)."

Our experience of the 'constant retinal disparity' condition appears to provide additional evidence for this conclusion. And this is something that those who favour (Johnston, 1991)'s account agree with me on, since her account predicts only a small effect of vergence on the scaling of disparities.

*4. "Fourth, stereo 'depth constancy' appears to be a cognitive (not perceptual) phenomenon, reflecting our experience of a world distorted in perceived stereo depth."*

This argument is developed in the third section (below), and best articulated there.

5. In my ECVP poster ([https://osf.io/6wn7x](https://osf.io/6wn7x)), I further argue that the 'Linton Stereo Illusion' highlights that we've been trying to hold two incompatible commitments at the same time:

a. On the one hand, the logic of stereo 'depth constancy' (even partial 'depth constancy') implies that observers are blind to the changes in the relative disparities of stimuli moving in depth.

b. On the other hand, if there's one thing we've know about stereo vision for well over a century (Howard, 1919), it's that observers are supremely sensitive to changes in relative disparities.

For me, the sensitivity to changes in the relative disparities seems to be the more fundamental of the two. So, it's no surprise that the 'Linton Stereo Illusion' appears to demonstrate that observers are not blind to changes in relative disparities in the way that 'depth constancy' suggests.

However, against my interpretation, two alternative explanations have been raised:

## 2. Alternative Explanation One: Failure of Depth Constancy?

A leading vision scientist has suggested to me in correspondence[4] that the 'Linton Stereo Illusion' is just an instance of a well-known phenomenon – the failure of stereo 'depth constancy' – and that the 'Linton Stereo Illusion' can be explained by the analysis in the leading paper in the field[5], (Johnston, 1991). By contrast, in this Section, I argue that (Johnston, 1991)'s account makes predictions that are not only inconsistent with, but the exact opposite, of what is observed in the 'Linton Stereo Illusion'.

I hope I can be forgiven for spending some time on (Johnston, 1991). As one reviewer notes, there are considerable dissimilarities between our stimuli, that I don't explore here. However, I would expect common principles of stereo vision to govern both cases. And I feel the field as a whole hasn't fully worked through the implications of (Johnston, 1991)'s account, meaning that they read the 'Linton Stereo Illusion' as supporting it (a common response I received at VSS and ECVP), when it does not. However, since a I don't want to turn this paper on the 'Linton Stereo Illusion' into a paper on (Johnston, 1991), I have provided an extended version of the following analysis in (Linton, 2024b).

Failure of stereo 'depth constancy' is a claim that dates back at least to (Helmholtz, 1866), who argued that distortions in stereo vision are due to visual system scaling disparities using a faulty estimate of the viewing distance (which is typically thought to come from vergence):

> "Owing to the uncertainty of our judgments as to the degree convergence of the eyes, we are liable to have illusions also about the forms of things in space seen binocularly. The interpretation of the visual phenomena would be correct if the amount of convergence were different, but it is not correct for the convergence actually used."

[4] Personal communication, 'There is no challenge to our understanding of stereo vision: Response to Linton and Kriegeskorte (ECVP 2024 and ArXiv, https://arxiv.org/abs/2408.00770)', 9th September 2024.

[5] As evidenced by most citations for an article on failure of stereo depth constancy (369 at time of writing).

(Helmholtz, 1866), pp.264-5 (trans. p.318).

For a representative sample of restatements of this over the last seven decades see (Wallach & Zuckerman, 1963) in the 1960s, (Ono & Comerford, 1977) in the 1970s, (Foley, 1980) in the 1980s, (Johnston, 1991) in the 1990s, (Scarfe & Hibbard, 2006) in the 2000s, (Volcic et al., 2013) in the 2010s, and (Yu et al., 2021) in the 2020s. It's fair to say (Johnston, 1991) has been the leading paper in the field for the last 33 years, and so it's understandable that this vision scientist focuses on it.

(Johnston, 1991) had participants adjust a stereo-defined cylinder so that it looked regular (dotted line in Figure 2). At medium distances (107cm), (Johnston, 1991) found close to veridical 3D shape perception. However, at close distances (53.5cm), a compressed cylinder looked regular, implying 'over-constancy' (accentuated depth) at near distances. Whilst at far distances (214cm), a cylinder accentuated in depth looked regular, implying 'under-constancy' (compressed depth) at far distances.

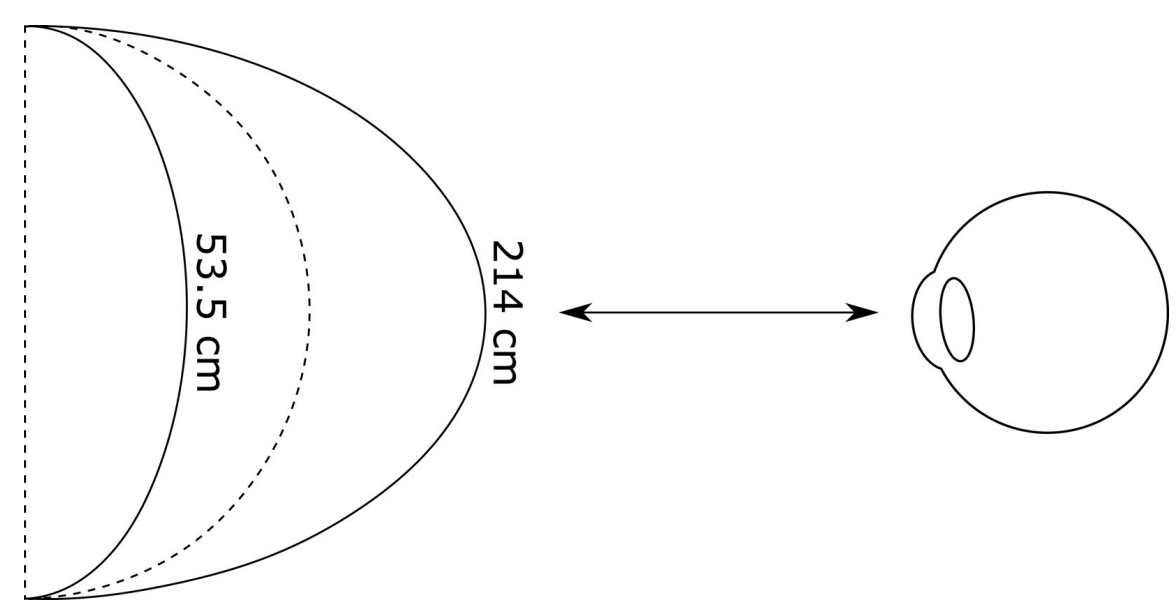

Figure 2: Disparity defined 3D shape of cylinders perceived to be 'regular' (dotted line) by observers in (Johnston, 1991) at near (53.5cm) and far (214cm) viewing distances.

(Johnston, 1991) concluded that binocular disparities are 'scaled' by the visual system using an inaccurate estimate ($y$) of the viewing distance ($x$), where:

$$y \approx \underline{80} + \underline{0.265*(x - 80)}$$

Weak Triangulation ↗     Strong Triangulation ↖

We can think of this formula as having two components:

1. Abathic Distance (Weak Triangulation): A default internal estimate of viewing distance, known as the 'abathic distance', that is roughly 80cm (at which distance $y = x$).

2. Vergence (Strong Triangulation): A refinement of the 'abathic distance' based on a compressed (gain of 0.265) estimate of the viewing distance, thought to come from 'vergence'.

I call these 'Weak Triangulation' and 'Strong Triangulation' because they are both classic triangulation accounts, the only difference is whether the disparities are being 'scaled' using a default internal estimate of the viewing distance (Weak Triangulation) or the visual system tries to estimate the viewing distance (Strong Triangulation). Both 'Weak Triangulation' and 'Strong Triangulation' imply a degree of 'depth constancy' (disparity scaling) for a stimulus moving in depth when vergence is fixed. And this, as we shall now see, gives us a way of differentiating my account from (Johnston, 1991)'s.

### a. ‘Constant Physical Separation’ Condition

You might think that (Johnston, 1991) provides an obvious explanation for our perceptual experience of the ‘constant physical separation’ condition, where we experience accentuated stereo depth as the stimulus moves towards us (Movie 2). Indeed, 20 years ago, (Scarfe & Hibbard, 2006) showed that if you move (Johnston, 1991)’s stereo cylinder in depth towards an observer, the cylinder is perceived as expanding in depth by the observer, noting:

> “A simple prediction from [(Johnston, 1991)] is that disparity-defined objects should appear to expand in depth when moving towards the observer, and compress when moving away.”

But the same stereo distortions are also predicted by my account (Linton, 2023), since the disparities are larger the closer the object is. So how can we differentiate between these two accounts? As we see in part c. (below) you can actually do it using (Johnston, 1991)’s original cylinder by alternating your fixation between the front and the back of the cylinder. But that’s not the approach the ‘Linton Stereo Illusion’ adopts. Instead, it introduces two key innovations:

First, differentiating the two accounts is easier when we focus on more complex stimuli. Specifically, if we have two sequential separations in depth (the red and blue arrows in Figure 3), we can explore how ‘distance scaling’ affects the perceived relative size of these two separations in depth.

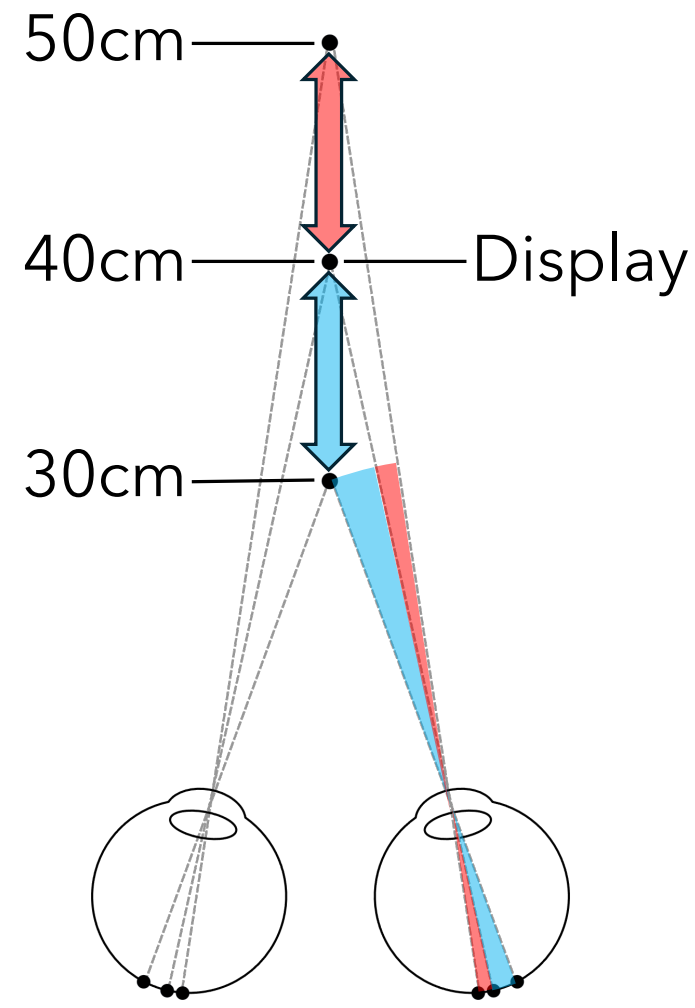

Figure 3. Viewing geometry in the ‘constant physical separation’ condition.

Second, we should pick a viewing distance that is *closer* than the ‘abathic distance’ (≈80cm), and here I use a viewing distance of 40cm. The reason is that both my (Linton, 2023) and (Johnston, 1991)’s accounts make similar predictions for stimuli that are *further* than the ‘abathic distance’ (>80cm), but they make the exact opposite predictions when the stimuli are *closer* than the ‘abathic distance’ (<80cm). (Johnston, 1991)’s account predicts that the stimulus in the ‘constant physical separation’ condition (when vergence is fixed at 40cm) should compress as it comes closer, whilst my account predicts that it should expand. So (Johnston, 1991) predicts the exact opposite pattern of distortions than we actually experience in the ‘constant physical separation’ condition (Movie 6).

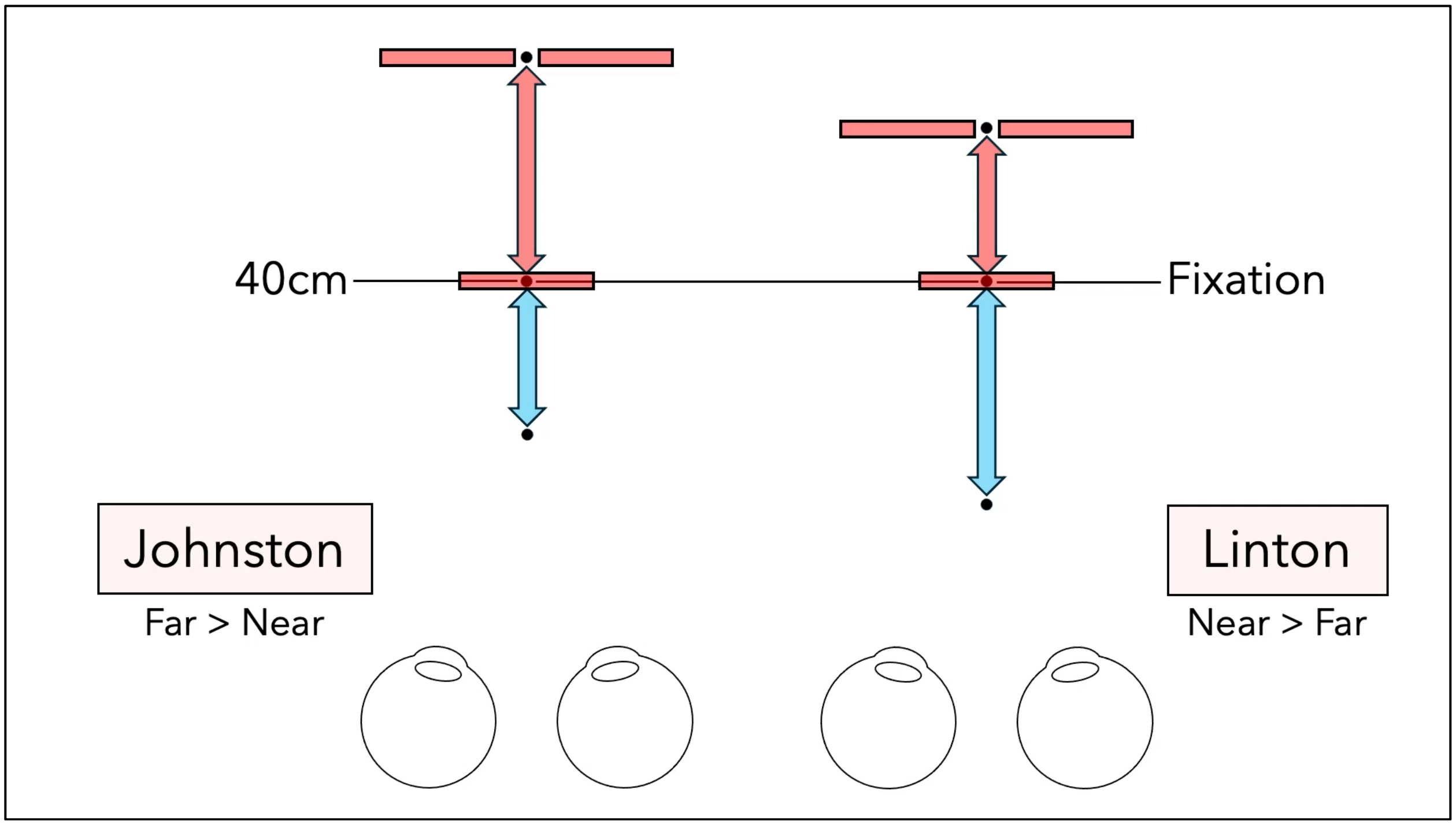

[Insert Movie 6 here: https://youtu.be/7chGl8JtPVE]

Movie 6. Perceived stereo distortions predicted by (Johnston, 1991) and (Linton, 2023) for the 'constant physical separation' condition of the 'Linton Stereo Illusion' (vergence fixed at 40cm).

This might seem counterintuitive, so let me explain why this follows from (Johnston, 1991)'s account. On (Johnston, 1991)'s account, the 'abathic distance' (mid-distance) acts as a 'flipping distance'. The stereo distortions of far viewing (2m+)(under-correction) gradually correct as we reach mid-distance (≈80cm) and then invert as we reach near viewing (40cm)(over-correction). In far viewing (2m+), (Johnston, 1991)'s account predicts progressive 'under-constancy' of equal separations in depth (blue arrow > red arrow). This reflects our experience of the real-world, where (for instance) the stereo depth between evenly spaced lampposts seems to get flatter and flatter with distance. But what this means is that in near viewing (40cm), (Johnston, 1991)'s account should predict the exact opposite: progressive 'over-constancy' of equal separations in depth (blue arrow < red arrow).

To understand why, consider how disparities fall off with the viewing distance squared. Imagine a simple case where the blue (near) and red (far) arrows have the same disparity (1 degree). Because disparity falls off drastically with the viewing distance squared, the further the fixation distance is, the larger the red (far) arrow has to be in order to produce the same disparity as the blue (near) arrow (Table 1). It therefore follows that if the visual system thinks the fixation distance is further than it actually is, it will disproportionately scale the red (far) arrow more than the blue (near) arrow.

| Vergence Angle | Vergence Distance | 1 Degree Disparity Near | 1 Degree Disparity Far | Ratio |
|---|---|---|---|---|
| 1 Degree | 183cm | 91cm | Horizon (∞) | 1:∞ |
| 2 Degree | 92cm | 31cm | 91cm | 1:3 |
| 3 Degree | 61cm | 15cm | 31cm | 1:2 |
| 4 Degree | 46cm | 9cm | 15cm | 1:1.7 |
| 5 Degree | 37cm | 7cm | 9cm | 1:1.3 |

Table 1. Lengths of the blue (near) arrow and red (far) arrow when they both correspond to 1 degree of disparity from fixation, for different fixation (vergence) distances. Note how ratio between the distances tends from 1: ∞ towards 1:1 as the fixation (vergence) distance reduces.

This is illustrated in Figure 4, which shows what happens when the disparities in our 'constant physical separation' condition (at 40cm) are scaled using progressively wrong estimates of the viewing distance up to, and including, (Johnston, 1991)'s own estimate of the viewing distance (70cm).

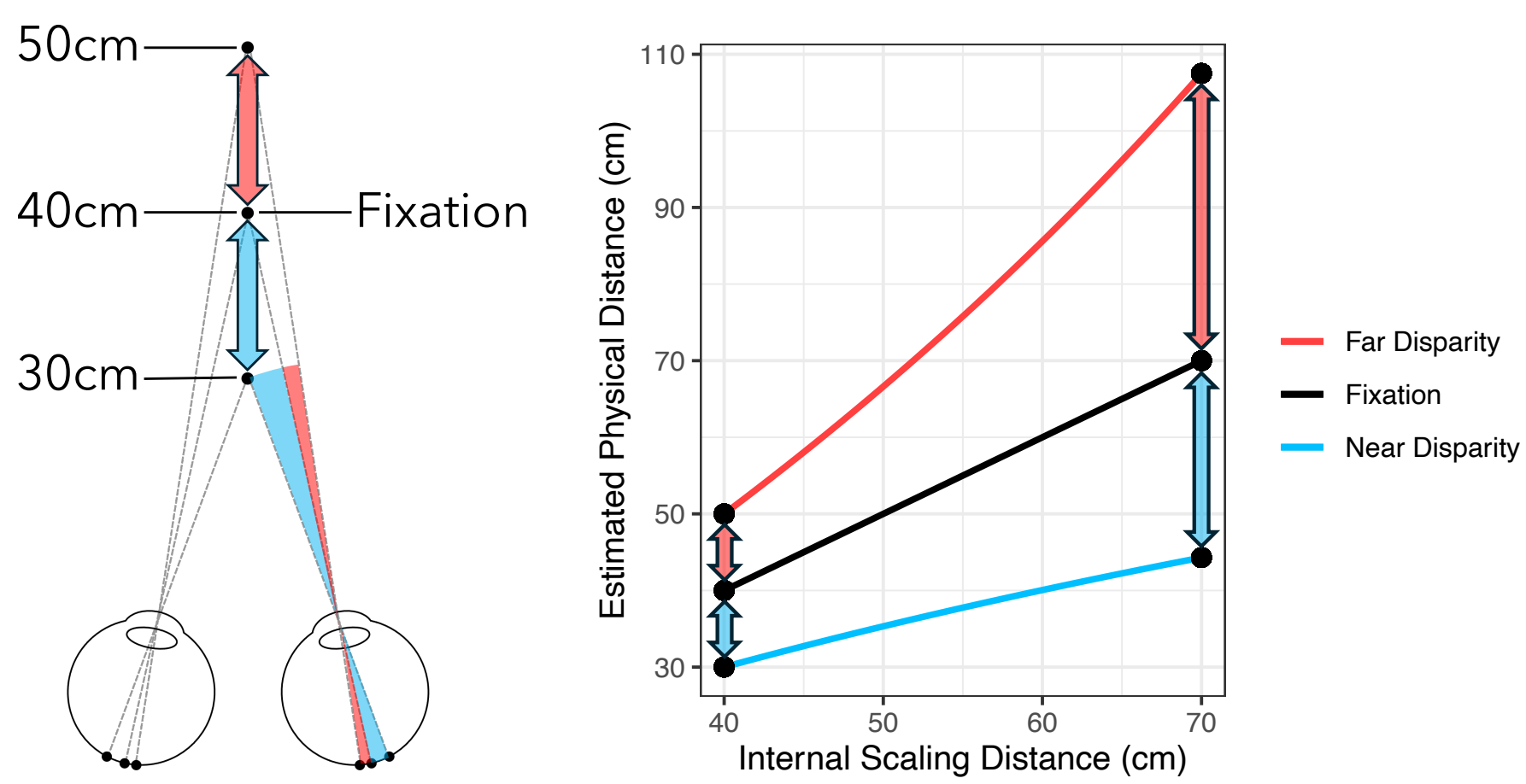

Figure 4. Disparities for the 'constant physical separation' condition of the 'Linton Stereo Illusion' replotted for different scaling distances (with vergence fixed at 40cm).

Now let me consider some concerns with this analysis of (Johnston, 1991):

First, the analysis in Figure 4 is prefaced on vergence remaining fixed at 40cm in the 'Linton Stereo Illusion'. You might object that vergence was not fixed in my VSS and ECVP demos. That is true. But we can re-test the illusion with fixed vergence. Set the screen at 40cm, fix your vergence on the screen, and watch the illusion. The separation between the circles appears to expand as it gets closer (as predicted by my account) not compress (as predicted by (Johnston, 1991)).

Second, you might object that (Johnston, 1991) simply got the 'scaling' ('abathic') distance wrong. On this account, the 'scaling' ('abathic') distance is closer than (Johnston, 1991) thought, but (Johnston, 1991)'s overall account still holds. But here's the dilemma. If you bring the 'scaling' distance closer – to, say, 40cm – then all you would predict is that the 'constant physical separation' condition should look undistorted as it moves in depth. To get the actual pattern of distortions predicted by my account (Movie 6), and actually observed in the 'constant physical separation' condition, the 'scaling' distance would have to be much closer than 40cm. But arguing that stereo vision is 'optimized' for viewing distances much closer than 40cm is hard to maintain, and the closest 'scaling' ('abathic') distances I have seen in the literature are 45cm in (Volcic et al., 2013) and 50cm in (Scarfe & Hibbard, 2006).

Further objections to trying to explain the 'constant physical separation' condition using (Johnston, 1991)'s account are discussed in (Linton, 2024b). Consider, for instance, the massive perceived absolute distances predicted by (Johnston, 1991)'s account for the two 10cm separations. It predicts the near 10cm separation should be perceived as 26cm, and the further 10cm separation should be perceived as 37cm, neither of which correspond to our actual experience of the illusion.

**b. 'Constant Retinal Disparity' Condition**

Next, let's model the 'constant retinal disparity' condition using (Johnston, 1991). The vision scientist argues that if vergence is tracking the stimulus, then (Johnston, 1991) would predict something close to the rigid percept we see in the 'constant retinal disparity' condition. So, let's model (Johnston, 1991)'s prediction with vergence tracking the front circle (moving from 40cm to 33cm), as they suggest:

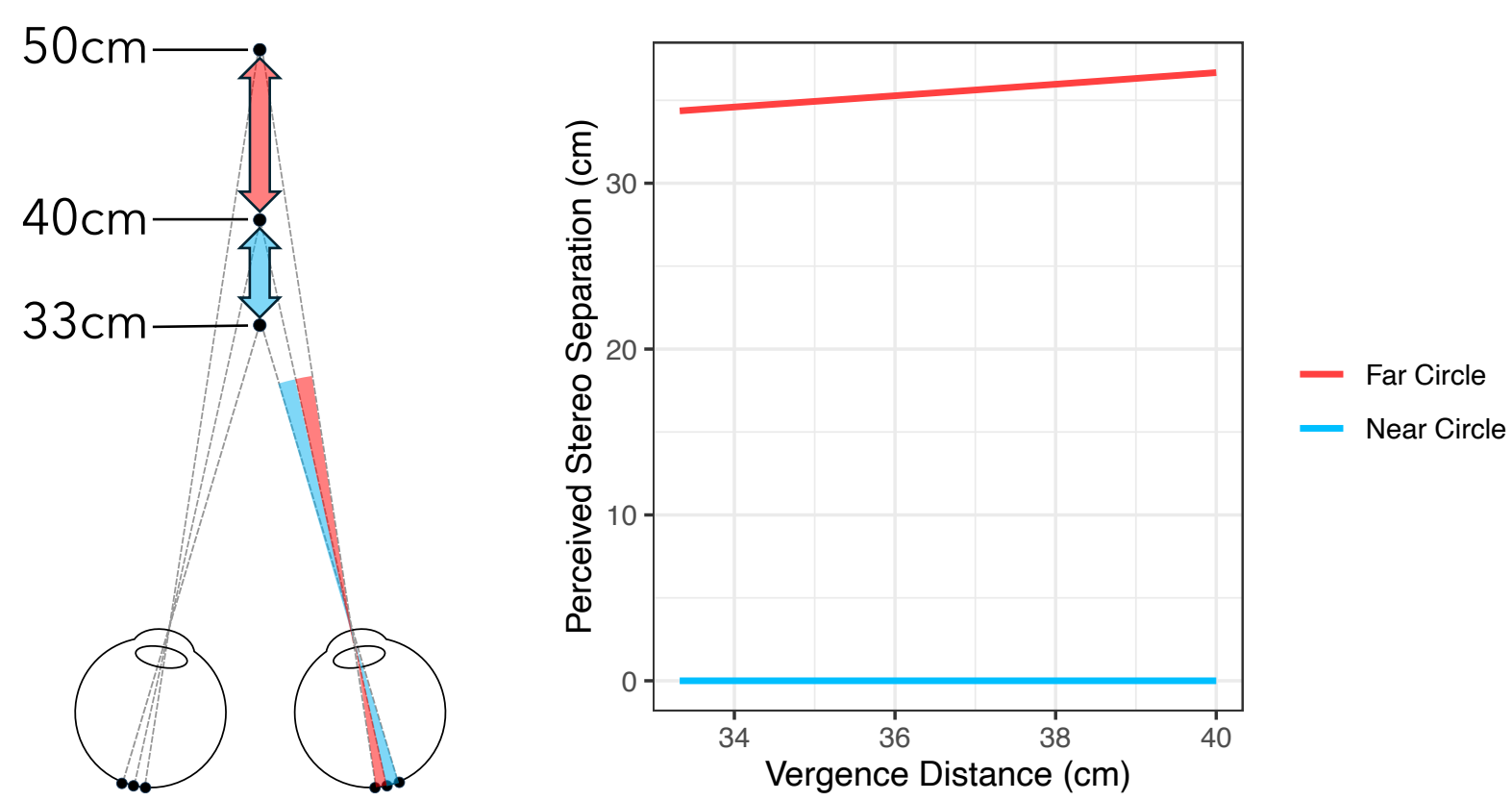

Figure 5. (Johnston, 1991)'s prediction of how the perceived stereo depth between the two circles should change as vergence tracks the near circle from 30cm to 40cm in the 'constant retinal disparity' condition of the 'Linton Stereo Illusion'.

You might look at Figure 5, and see the lines are roughly parallel, and therefore conclude that (Johnston, 1991)'s account predicts something close to my account (that the near and far circle move rigidly together). But here's the problem. The lines are not parallel, so the near and far circles are not predicted to move rigidly in depth together. Instead, Figure 5 predicts that we should see the separation between circles compress by 2.3cm as the circles come towards us. This is almost as big as the 3.3cm that the circles actually physically compress to keep disparity constant. So (Johnston, 1991)'s account predicts that we should perceive something close to physical reality, not retinal disparities.

The problem is that the 2.3cm compression in Figure 5 is masked by the implausibly large absolute distances that (Johnston, 1991)'s account predicts the separation is scaled to. According to (Johnston, 1991), the 6.7cm–10cm separation should be perceived as 34.4cm–36.7cm. But our experience in no way coheres with this. So masking the 2.3cm compression predicted by (Johnston, 1991)'s account comes at the cost of accepting other implausible predictions about absolute distance.

**c. Stimulus at Starting Position**

One of the shortcomings of (Johnston, 1991)'s account, that has so far gone unnoticed by the literature, is that it predicts that we should experience massive stereo distortions as we look around the scene. This is because whatever the actual vergence distance is, the disparities in the scene will always be rescaled as if the vergence distance were the 'abathic' distance (≈80cm).

Let's illustrate this point with the 'Linton Stereo Illusion' in its starting position (which is common for both conditions): a front circle at 40cm, and a back circle at 50cm. On (Johnston, 1991)'s account, when I fixate on the back circle (50cm), the visual system thinks vergence is ≈80cm. And when I fixate on the front circle (40cm), the visual system thinks vergence is at ≈80cm. But remember that the visual system scales disparities very differently depending on whether they are in front of fixation or behind fixation (Table 1). So, when I fixate on the back circle (50cm), the 10cm separation in front of fixation is scaled to 19cm. But when I fixate on the near circle (40cm), the 10cm separation behind fixation is scaled to 37cm. Consequently, (Johnston, 1991)'s account predicts that the perceived separation between the front and back circles should expand by ≈50% as we shift our fixation from the back circle (50cm) to the front circle (40cm). But this does not cohere with our experience.

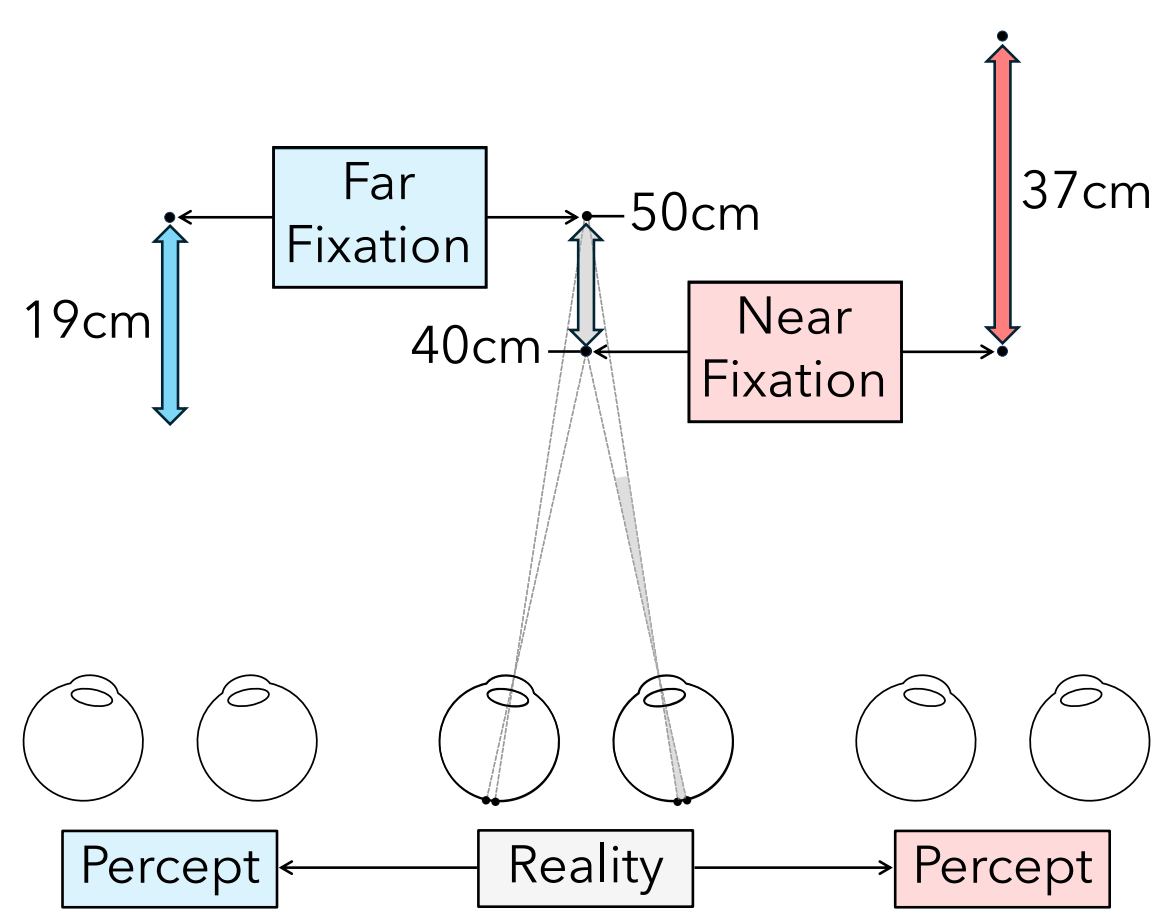

Figure 6. Illustration of how the perception of the very same physical distance distorts with near and far fixation under (Johnston, 1991)'s account. (Johnston, 1991)'s account predicts that our stereo depth percept should expand by ≈50% as we shift our fixation from far (50cm) to near (40cm).

Instead, our stereo percept seems pretty stable (doesn't distort) as we shift our fixation from front (50cm) to back (40cm). Ironically, this is the one thing that *full*-'depth constancy' accounts (Guan & Banks, 2016) and my 'minimal' (disparity-only) model (Linton, 2023) agree on, namely that our stereo depth percept is largely[6] invariant to eye movements, either because there is no re-scaling of disparities with each eye movement (my account), or because the re-scaling of disparities with each eye movement is veridical (Guan & Banks, 2016). By contrast, on (Johnston, 1991)'s *partial*-'depth constancy' account, the rescaling of disparities with each eye movement is not veridical, leading to the prediction of massive distortions of stereo depth as we look around the scene.

[6] There will be some very subtle changes to the disparities on the retina with vergence eye movements ('gaze contingent disparities': (Turski, 2016), (Linton, 2019)) due to the slight offset of the nodal point and the centre of rotation of the eye. But the key point is that retinal disparities are not being re-scaled as having different perceived depths.

**Alternative Explanation Two: Artefact of Constant Angular Size?**

When an object moves towards us in the real world, its 'angular size' (size on the retina) expands. But in the 'Linton Stereo Illusion' I keep the 'angular size' of the stimulus constant. This is a standard procedure to ensure we are testing motion in depth from stereo alone, rather than from another cue such as 'looming' (the changing 'angular size' as the stimulus moves towards us).

But an alternative interpretation is that the 'Linton Stereo Illusion' is simply an artefact of keeping 'angular size' constant. There are two reasons to take this concern seriously. First, we don't notice objects distorting in depth in normal viewing conditions. Second, if we re-render the 'Linton Stereo Illusion' so angular size does change realistically, the distortions in the 'Linton Stereo Illusion' are no longer apparent. If we simulate everyday vision by adding angular size changes to the 'constant physical separation' condition, the stimulus now appears to move rigidly (Movie 7, right).

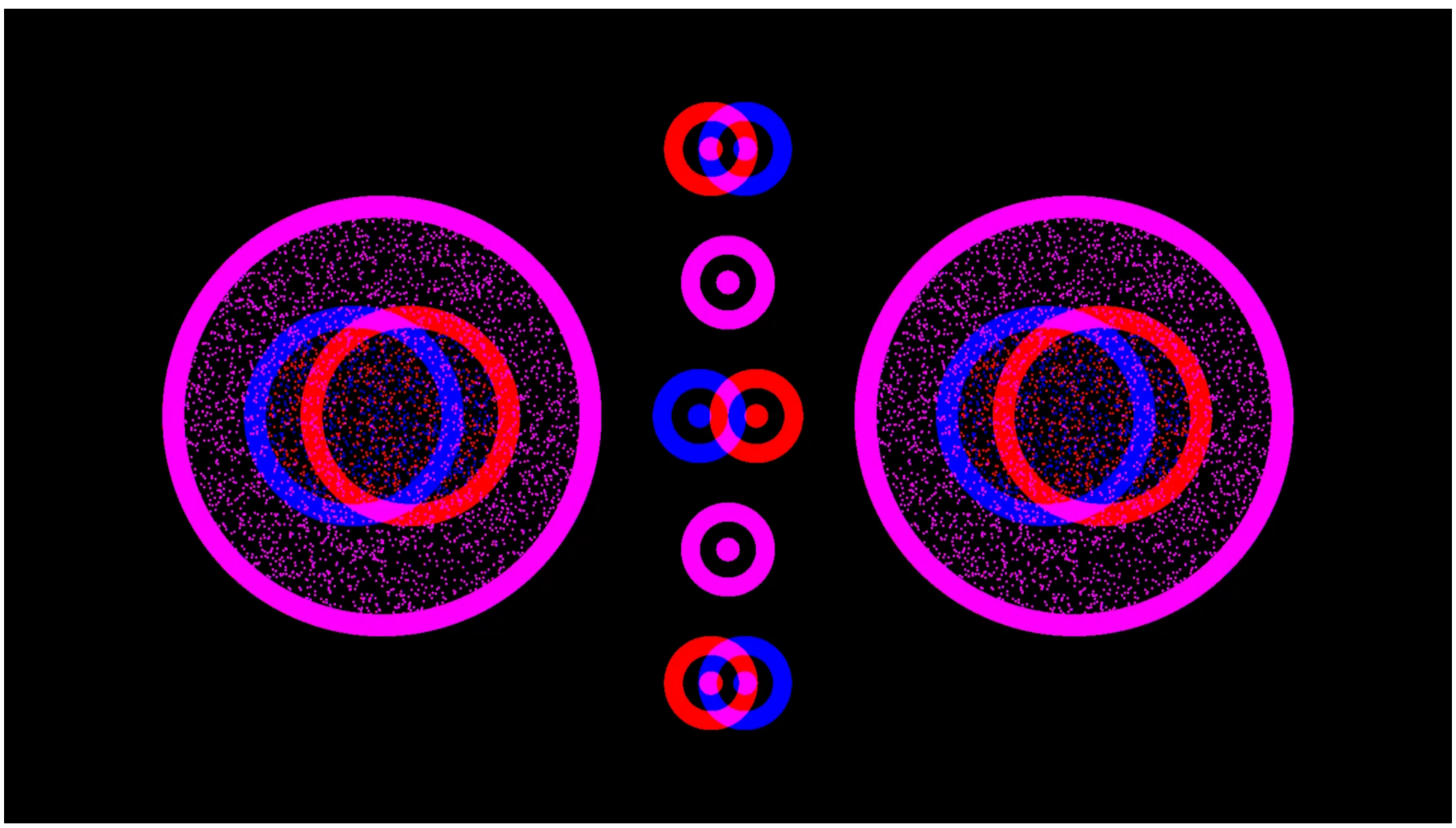

[Insert Movie 7 here: https://youtu.be/PzpFXTAm13Q]

Movie 7. 'Linton Stereo Illusion' Demo 3: 'Constant physical separation' condition with angular size kept fixed (left) or veridically changing with distance (right). In the middle are three static stereo targets staggered at 30cm (in front of screen), 40cm (on screen), and 50cm (behind screen). Stimuli specified for a 27-inch display at a 40cm viewing distance with 6.4cm interpupillary distance. Demos available at: https://Github.com/StereoIllusion/

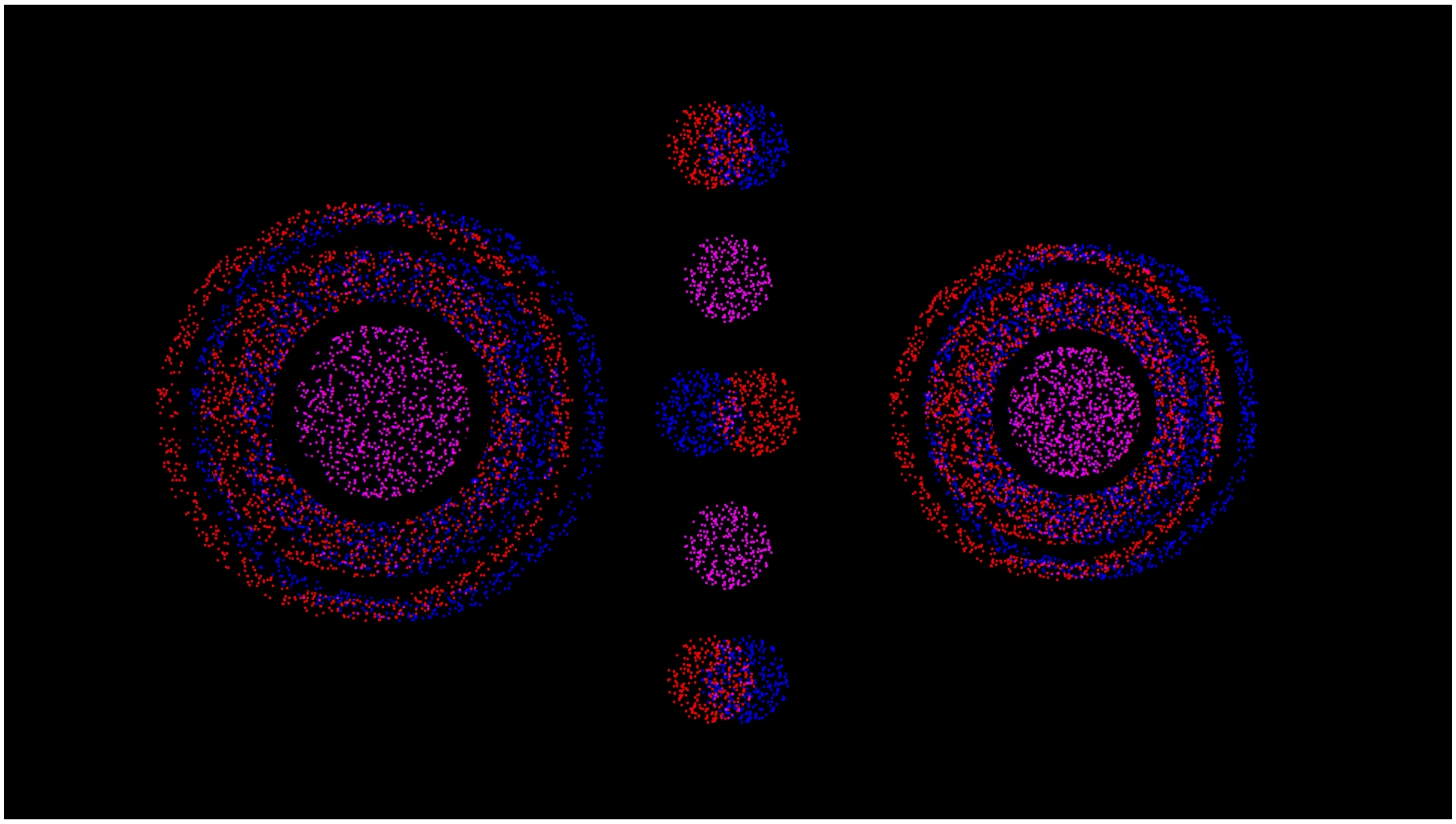

[Insert Movie 8 here: https://youtu.be/BqgLcCZVD8Q]

Movie 8. 'Linton Stereo Illusion' Demo 4: 'Constant physical separation' condition with angular size kept fixed (left) or veridically changing with distance (right), optimized for red-blue glasses. In the middle are three static stereo targets staggered at 30cm (in front of screen), 40cm (on screen), and 50cm (behind screen). Stimuli specified for a 27-inch display at a 40cm viewing distance with 6.4cm interpupillary distance. Demos available at: https://Github.com/StereoIllusion/

By in large, I agree with this assessment. But I wouldn't see it as a criticism. Instead, I think it shows that the 'Linton Stereo Illusion' tells us something interesting about (a) stereo vision, (b) cue integration, and (c) 'depth constancy', even if we don't notice its effects in everyday viewing.

**a. Stereo Vision**

The objection is that, by keeping angular size constant, we introduce a 'cue-conflict'. Stereo vision says the stimulus is moving in depth, but the absence of 'looming' says it is not, leading (on traditional cue integration accounts) to a reduction in the perceived motion in stereo depth (Landy et al., 1995).

There are three responses to this objection:

First, we can test whether perceived motion in stereo depth really is reduced in the 'fixed angular size' condition, and show that this is not true. The obvious test of whether keeping 'angular size' fixed leads to a reduction in perceived stereo motion-in-depth, is to place static stereo targets at the distances the motion should hit, and see if the stimulus in fact hits these targets or falls short. Look at the motion of the 'fixed-angular size' stimulus in Movie 7 (left). You can see that it doesn't fall short of the 30cm or 50cm static stereo targets in the middle of the display. Indeed, the original version of the 'Linton Stereo Illusion'[7] had static targets at 30cm, 40cm, and 50cm, to address this very concern.

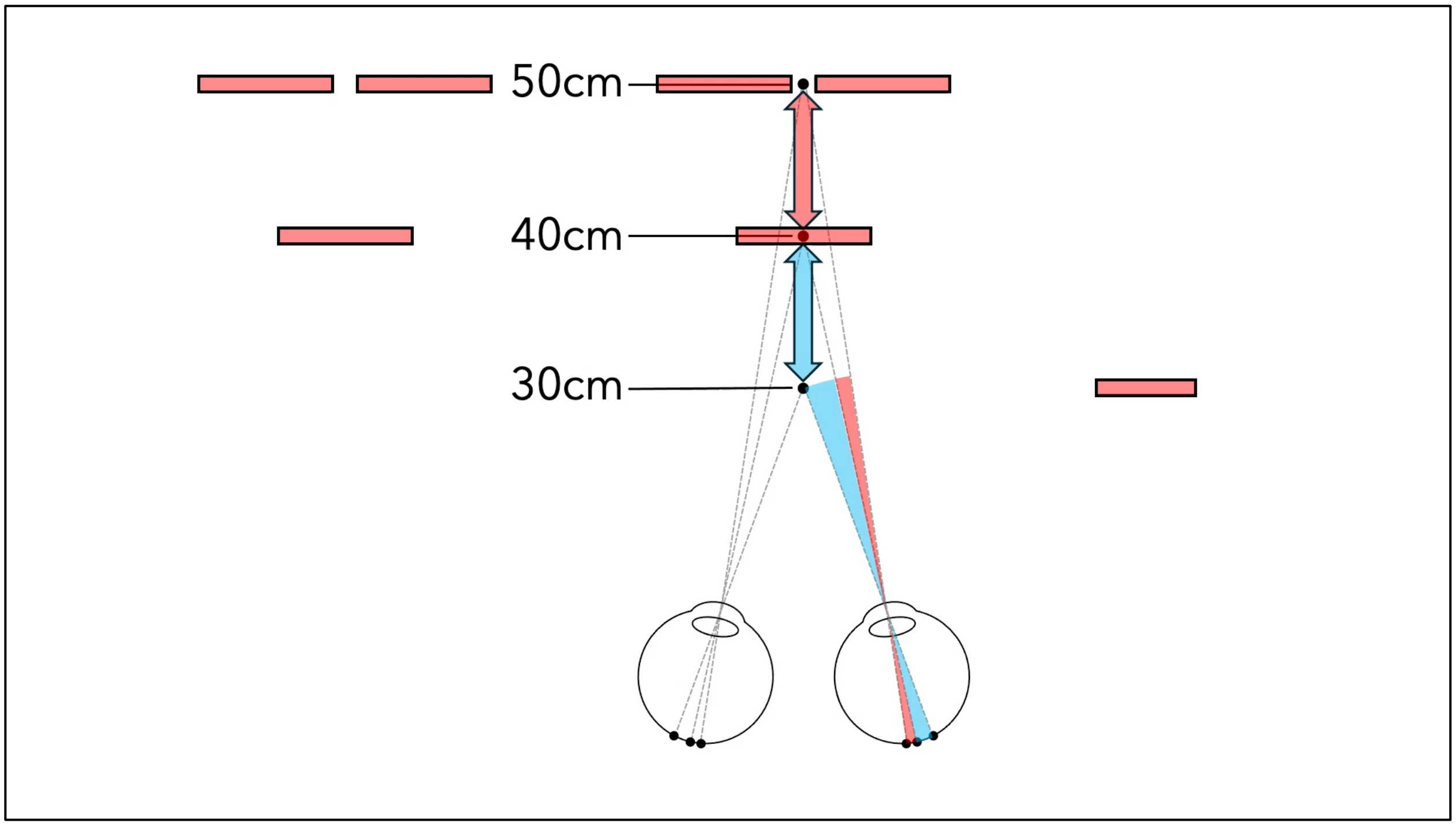

[Insert Movie 9 here: https://youtu.be/xWAZvpJkWUY]

Movie 9. Original version of the 'Linton Stereo Illusion' submitted to VSS Demo Night 2024.

One reviewer suggests that they don't see the back circle in the fixed angular size stimulus moving as far back as the back circle in the changing angular size stimulus. This is something to explore with additional manipulations of the stimuli in Demos 3 and 4 (e.g. additional stereo targets at 50cm).

Second, the 'Linton Stereo Illusion' still appears to work even if we don't control for angular size, so long as we use dots as the stimulus. This seems like an obvious solution, since the dots' angular size doesn't change that much with distance. But the problem here is diplopia. I have a convincing demo in the lab – the motion of the dots in front of the screen looks much closer to the motion of the dots behind the screen in the 'constant retinal disparity' condition than in the 'constant physical separation' condition – but the anaglyph version[8] does not fully capture this effect.

Third, we can turn the challenge back on 'cue conflict' account. How can it explain our experience of the original 'constant retinal disparity' condition (Movie 3) as rigid?

In the 'constant retinal disparity' condition, a stimulus with constant retinal disparity is seen as moving rigidly in depth. On my account, this is because constant disparities are perceived as having constant stereo depth. But the 'cue conflict' account can't appeal to this explanation. Indeed, it's the very thing they're rejecting. Otherwise, they wouldn't be objecting to my account. So instead, they must claim two things: First, that constant disparities are not perceived as having constant stereo depth for a stimulus moving in depth. Instead, a stimulus with constant disparities should be seen as compressing in depth as the stimulus comes closer. Second, and this is the difficult part, they must argue that the 'cue conflict' introduced by keeping 'angular size' fixed just so happens to cancel out the compression in perceived stereo depth *by exactly the right amount* so that we don't see the stimulus as compressing, or as expanding, but as rigid as it moves in depth. But why should these two depth cues – stereo and fixed 'angular size' – cancel each other out so perfectly? On the 'cue conflict' account, this is entirely a coincidence, and nothing to do with constant disparities having constant stereo depth.

[7] Version submitted to VSS Demo Night (17th March 2024): https://www.youtube.com/watch?v=El7YKaURsuQ
[8] Demo 5 / Movie 10. https://Github.com/StereoIllusion/

The problem is that our experience of the 'constant retinal disparity' condition is exactly as my 'minimal' (disparity-only) account predicts. So, advocates of the alternative 'cue conflict' explanation have to explain why their account mimics my account, whilst also explicitly rejecting my account.

This discussion of stereo vision should be enough to convince you that the 'Linton Stereo Illusion' can't be written off simply as an artefact of keeping 'angular size' constant. Indeed, our leading theories of stereo vision struggle to explain why such an 'artefact' (of a 'constant retinal disparity' stimulus appearing to have constant stereo depth) should arise in the first place.

In the remainder of the paper, I consider the implications for 'cue integration' and 'depth constancy' that appear to arise from the 'Linton Stereo Illusion', but these are less essential to the core message.

**b. Cue Integration**

It's typically assumed that 3D vision integrates the various different depth cues into a single coherent percept at the level of our visual experience ('mandatory fusion')(Hillis et al., 2002). In which case, one tempting conclusion about the 'Linton Stereo Illusion' might be that whilst it teaches us something important about stereo vision, it doesn't really teach us anything interesting about 3D vision in normal viewing conditions. On this account, multiple depth cues bring us closer to the veridical percept, and the depth distortions inherent in stereo vision – whether they are (Johnston, 1991)'s or (Linton, 2023)'s – disappear (Johnston et al., 1994; Scarfe & Hibbard, 2013). This explains why, when we add changes in 'angular size' to the 'constant physical separation' condition, the distortions suddenly disappear.

On this account, adding or removing the changing 'angular size' cue leads to a different 3D percept. Whilst this interpretation of the 'Linton Stereo Illusion' might be tempting, I don't believe it is correct. The problem is that it predicts that the perceived stereo depth (i.e. the perceived real-world depth) of the stimulus is changed by the addition of the 'angular size' cue. But, as I have just argued, this doesn't seem to be the case. The obvious test (as we saw in Movie 7) is to assess the motion-in-depth of the '*changing*-angular size' stimulus against static stereo targets at 30cm, 40cm, and 50cm. Its motion-in-depth matches the static stereo targets. But so, too, does the motion-in-depth of the '*fixed*-angular size' stimulus. Leading us to the conclusion that they must both have the same perceived motion-in-depth, even if we can't directly judge this to be the case.

Now I agree that our *judgements* of the perceived motion-in-depth of the two stimuli in Movie 7 are different. One challenge for my account is to explain why this is. But it's also a challenge for 'cue integration' accounts, because if it were simply a case of the two stimuli have different *perceived* motions-in-depth, the static stereo targets wouldn't be a good match for both stimuli.

So, my explanation is that whilst both 'stereo' and 'looming' (changes in 'angular size') affect our *judgement* of perceived motion-in-depth, only stereo actually affects our *perception* of motion-in-depth. The argument, from (Linton, 2023), (Linton, 2017), Ch2, has two stages:

First, I treat 'looming' as a 'pictorial cue'. 'Looming' is only a cue to 'depicted' (pictorial) depth, not *perceived* depth. For instance, when an object 'looms' (comes towards you) on TV, it doesn't come out of the screen in perceived real-world depth. This is because the perceived real-world depth in this scenario is dictated by stereo vision alone. I argue that the same principle applies here, too.

Second, 'pictorial cues' induce a 'cognitive bias' that inhibits our ability to accurately introspect on our own visual experience. In our judgments of our perception, we (cognitively) mistake '*depicted*' motion-in-depth (from the 'looming cue') for *perceived* real-world motion-in-depth. But our visual system is not so easily fooled. It produces *perceived* real-world depth on the basis of stereo vision alone (explaining why both stimuli in Movie 7 have the same perceived depth relative to the static stereo targets).

So the '*depicted*' motion-in-depth (from the 'looming cue') is a 'cognitive bias' that we cannot *directly* overcome when we introspect on our own visual experience ((Linton, 2017), Ch2; (Linton, 2023)). We can only overcome this failure of introspection *indirectly* through 'structured introspection': 'scaffolding' visual space with static stereo targets to help us judge which depth cues are contributing to perceived real-world depth, and which ones are merely contributing to 'depicted' depth.

This requires a strong modularity thesis between 'perception' and 'cognition' ((Linton, 2017), Ch2; (Linton, 2023)): the idea that whilst we are (cognitively) fooled, our visual systems are not. Strong modularity is not necessarily an uncommon approach in the literature, but usually runs in the opposite direction: the idea that whilst we are not (cognitively) fooled, our visual systems are (for instance, by the 'Müller-Lyer illusion'). And, as (Block, 2023) notes, there is a lively debate where to draw the line:

> "My usage of the terms "cognition" and "perception" is consonant with much of the recent literature on perception ((Firestone & Scholl, 2016); (Pylyshyn, 1999)), but some restrict the term "perception" to what I am calling low-level perception (Linton, 2017) and others use "cognition" to encompass mid-level and high-level perception as both perception and cognition (Cavanagh, 2011)."
>
> (Block, 2023), p.13.

The point of this discussion isn't to argue the 'Linton Stereo Illusion' in Movie 7 vindicates my account, only to highlight the dilemma it appears to pose for traditional 'cue-integration' accounts. Why are both stimuli *perceived* to move veridically relative to the static stereo targets if, as the "cue-integration' account suggests, the difference in the *judged* motion-in-depth of the stimuli in Movie 7 is attributed to the fact they have different *perceived* motions-in-depth?

**c. Depth Constancy**

On my account, we experience a world distorted in perceived stereo depth, reflecting the distorted disparities on our retina. But we typically don't notice these distortions in everyday viewing.

One reason is that disparities at far distances are relatively small. So, we shouldn't expect to notice a car distorting in perceived stereo depth as it drives towards us. But this isn't the complete answer, as these distortions are very significant at near viewing distances.

Another reason, as we have just seen, is that changes in 'angular size' can (cognitively) mask these distortions (Movie 7). There might be good reasons for cognitive processes relying on pictorial cues, since pictorial cues are much more stable[9] with changes in viewing distance than stereo vision is.

Another reason we don't notice these distortions in everyday viewing is that what we really care about is 'shape constancy' not 'depth constancy'. 'Depth constancy' focuses on the distortion of z-axis points with distance. Since disparities fall off with the viewing distance squared, we would expect z-axis depth to be distorted in the same way. But what people really care about is 'shape constancy': whether objects look distorted in 3D shape with distance. This focuses on the relationship of z-axis depth to x-axis width and y-axis height. Since the x-axis width and y-axis height of objects fall-off linearly with distance due to perspective, full perceptual z-axis 'depth constancy' would look weird: a 3D object would appear to expand in z-axis depth (relative to its x-axis width and y-axis height) as it recedes. Conversely, the linear fall-off of x-axis width and y-axis height with distance makes the z-axis fall-off of depth with distance squared appear to be merely a linear fall-off of z-axis depth. And this is much closer to our actual everyday visual experience, where objects do seem to get flatter in stereo depth with distance.

**Additional Resources**

'Linton Stereo Illusion' Webpage: https://StereoIllusion.Github.io
'Linton Stereo Illusion' GitHub: https://Github.com/StereoIllusion/

'Five Illusions' Webpage: https://FiveIllusions.Github.io/
'Five Illusions' GitHub: https://Github.com/FiveIllusions

**Acknowledgements**

This research project and related results were made possible by the support of the NOMIS Foundation. This research was conducted in Nikolaus Kriegeskorte's Visual Inference Lab at Columbia University's Zuckerman Mind Brain Behavior Institute, with support from the NOMIS Foundation ('New Theory of Visual Experience' grant to PL at the Italian Academy for Advanced Studies, Columbia University), support from the Italian Academy for Advanced Studies, Columbia University, and support from the Presidential Scholars in Society and Neuroscience (PSSN), Columbia University. I thank Prof. Kriegeskorte, as well as David Freedberg and Elena Aprile (at the Italian Academy), and Christopher Peacocke, Pamela Smith, and Carol Mason (at PSSN). I thank audiences at Rochester Institute of Technology and the Center for Visual Science, University of Rochester (February 2024), Applied Vision Association Spring Meeting (March 2024), Royal Holloway, University of London (March 2024), Vision Sciences Society Demo Night (May 2024), and European Conference on Visual Perception and ECVP Illusion Night (August 2024) where this work was presented, as well as the lively discussion on X.com (Twitter) with 16,500 views:

https://x.com/LintonVision/status/1788572440034947469

This work was also presented as one of "Five Visual Illusions Challenge Our Understanding of Visual Experience" (Linton, 2025) at Vision Sciences Society (2025) and VSS Demo Night (2025), European

[9] Although not perfectly stable, see 'perspective distortions' (Cooper et al., 2012).

Conference on Visual Perception (2025) and ECVP Demo Night (2025), Optica Fall Vision Meeting (2025), Applied Vision Association (2025), and Society for Philosophy and Neuroscience (2025).

I thank the two reviewers for their very helpful comments on the manuscript. And thanks to Rob Allison, Marty Banks, Randolph Blake, Marisa Carrasco, Patrick Cavanagh, Aniruddha Das, Deborah Giaschi, Hany Farid, Aaron Hertzmann, Paul Hibbard, Patrick Hughes, Anya Hurlbert, Mike Landy, Ken Nakayama, Andrew Parker, Jenny Read, Austin Roorda, Mike Shadlen, Shin Shimojo, Christopher Tyler, Xiaoye (Michael) Wang, Laurie Wilcox, Niall Williams, Qasim Zaidi, and Li Zhaoping, for helpful discussions at VSS, ECVP, Optica FVM, and Columbia University.

I thank Phillip Guan and Olivier Mercier, Peter Giokaris, Zenna Tavares and Shaiyan Keshvari, Eivinas Butkus, and Guy Liotto for discussions on stereo displays / operating systems.

**References**

Adelson, T. E. (1995). *Checker shadow illusion: Http://persci.mit.edu/gallery/checkershadow*.

Banks, M. (2000). Viewing Geometry and Stereoscopic Vision. *NSF Award Abstract #9983387*. https://www.nsf.gov/awardsearch/showAward?AWD_ID=9983387

Block, N. (2023). *The Border Between Seeing and Thinking*. Oxford University Press.

Cavanagh, P. (2011). Visual cognition. *Vision Research*, *51*(13), 1538–1551. https://doi.org/10.1016/j.visres.2011.01.015

Cooper, E. A., Piazza, E. A., & Banks, M. S. (2012). The perceptual basis of common photographic practice. *Journal of Vision*, *12*(5), 8. https://doi.org/10.1167/12.5.8

Descartes, R. (1637). *Dioptrique (Optics). In J. Cottingham, R. Stoothoff, & D. Murdoch (Eds.), The Philosophical Writings of Descartes: Volume 1 (1985)*. Cambridge University Press.

Firestone, C., & Scholl, B. J. (2016). Cognition does not affect perception: Evaluating the evidence for "top-down" effects. *The Behavioral and Brain Sciences*, *39*, e229. https://doi.org/10.1017/S0140525X15000965

Foley, J. (1980). Binocular Distance Perception. *Psychological Review*, *87*, 411–434. https://doi.org/10.1037//0033-295X.87.5.411

Gregory, R. L. (1970). *The intelligent eye*. Weidenfeld and Nicolson.

Guan, P., & Banks, M. S. (2016). Stereoscopic depth constancy. *Phil. Trans. R. Soc. B*, *371*(1697), 20150253. https://doi.org/10.1098/rstb.2015.0253

Helmholtz, H. (1866). *Handbuch der Physiologischen Optik, Vol.III (translated by J. P. C. Southall 1925 Opt. Soc. Am. Section 26, reprinted New York: Dover, 1962).*

Hillis, J. M., Ernst, M. O., Banks, M. S., & Landy, M. S. (2002). Combining Sensory Information: Mandatory Fusion Within, but Not Between, Senses. *Science*, *298*(5598), 1627–1630. https://doi.org/10.1126/science.1075396

Howard, H. J. (1919). A Test for the Judgment of Distance. *Transactions of the American Ophthalmological Society*, *17*, 195–235.

Johnston, E. B. (1991). Systematic distortions of shape from stereopsis. *Vision Research*, *31*(7), 1351–1360. https://doi.org/10.1016/0042-6989(91)90056-B

Johnston, E. B., Cumming, B. G., & Landy, M. S. (1994). Integration of stereopsis and motion shape cues. *Vision Research*, *34*(17), 2259–2275. https://doi.org/10.1016/0042-6989(94)90106-6

Julesz, B. (1960). Binocular depth perception of computer-generated patterns. *Bell System Technical Journal*, *39*, 1125–1162. https://doi.org/10.1002/j.1538-7305.1960.tb03954.x

Kepler, J. (1604). Paralipomena to Witelo (W. H. Donahue, Trans.). In *Optics: Paralipomena to Witelo and Optical Part of Astronomy.* Green Lion Press, 2000.

Kleiner, M., Brainard, D., & Pelli, D. (2007). Kleiner M, Brainard D, Pelli D, 2007, "What's new in Psychtoolbox-3?" Perception 36 ECVP Abstract Supplement. *Perception*, *36*(ECVP Abstract Supplement).

Landy, M. S., Maloney, L. T., Johnston, E. B., & Young, M. (1995). Measurement and modeling of depth cue combination: In defense of weak fusion. *Vision Research*, *35*(3), 389–412.

Linton, P. (2017). *The Perception and Cognition of Visual Space*. Palgrave Macmillan.

Linton, P. (2019). *Would Gaze-Contingent Rendering Improve Depth Perception in Virtual and Augmented Reality?* https://arxiv.org/abs/1905.10366v1

Linton, P. (2020). Does vision extract absolute distance from vergence? *Attention, Perception, & Psychophysics*, *82*(6), 3176–3195. https://doi.org/10.3758/s13414-020-02006-1

Linton, P. (2021a). Does Vergence Affect Perceived Size? *Vision*, *5*(3), Article 3. https://doi.org/10.3390/vision5030033

Linton, P. (2021b). V1 as an egocentric cognitive map. *Neuroscience of Consciousness*, *7*(2), 1–19. https://doi.org/10.1093/nc/niab017

Linton, P. (2023). Minimal theory of 3D vision: New approach to visual scale and visual shape. *Philosophical Transactions of the Royal Society B: Biological Sciences*, *378*(1869), 20210455. https://doi.org/10.1098/rstb.2021.0455

Linton, P. (2024a). *Linton Stereo Illusion* (arXiv:2408.00770). arXiv. https://doi.org/10.48550/arXiv.2408.00770

Linton, P. (2024b). *Linton Stereo Illusion: Response on Johnston (1991)* (arXiv:2410.19740). arXiv. https://doi.org/10.48550/arXiv.2410.19740

Linton, P. (2025). *Five Illusions Challenge Our Understanding of Visual Experience*. Cognitive Computational Neuroscience. https://doi.org/10.5281/zenodo.17566980

Linton, P., & Kriegeskorte, N. (2024a). *Perceived Stereo Depth reflects Retinal Disparities, not 3D Geometry*. Applied Vision Association 2024 Spring Meeting. https://doi.org/10.5281/zenodo.17666523

Linton, P., & Kriegeskorte, N. (2024b). *Perceived Stereo Depth reflects Retinal Disparities, not 3D Geometry*. European Conference on Visual Perception. https://doi.org/10.5281/zenodo.17666724

Ono, H., & Comerford, J. (1977). Stereoscopic depth constancy. In W. Epstein (Ed.), *Stability and constancy in visual perception: Mechanisms and process*. Wiley.

Popper, K. R. (1963). *Conjectures and Refutations: The Growth of Scientific Knowledge*. Routledge & K. Paul.

Pylyshyn, Z. (1999). Is vision continuous with cognition?: The case for cognitive impenetrability of visual perception. *Behavioral and Brain Sciences*, *22*(3), 341–365. https://doi.org/10.1017/S0140525X99002022

Rogers, B. (2023). When is a disparity not a disparity? Toward an old theory of three-dimensional vision. *I-Perception*, *14*(5), 20416695231202726. https://doi.org/10.1177/20416695231202726

Scarfe, P., & Hibbard, P. B. (2006). Disparity-defined objects moving in depth do not elicit three-dimensional shape constancy. *Vision Research*, *46*(10), 1599–1610. https://doi.org/10.1016/j.visres.2005.11.002

Scarfe, P., & Hibbard, P. B. (2013). Reverse correlation reveals how observers sample visual information when estimating three-dimensional shape. *Vision Research*, *86*, 115–127. https://doi.org/10.1016/j.visres.2013.04.016

Turski, J. (2016). On binocular vision: The geometric horopter and Cyclopean eye. *Vision Research*, *119*, 73–81. https://doi.org/10.1016/j.visres.2015.11.001

Volcic, R., Fantoni, C., Caudek, C., Assad, J. A., & Domini, F. (2013). Visuomotor Adaptation Changes Stereoscopic Depth Perception and Tactile Discrimination. *Journal of Neuroscience*, *33*(43), 17081–17088. https://doi.org/10.1523/JNEUROSCI.2936-13.2013

Wallach, H., & Zuckerman, C. (1963). The constancy of stereoscopic depth. *The American Journal of Psychology*, *76*, 404–412.

Wheatstone, C. (1838). Contributions to the physiology of vision. —Part the first. On some remarkable, and hitherto unobserved, phenomena of binocular vision. *Philosophical Transactions of the Royal Society of London*, *128*, 371–394. https://doi.org/10.1098/rstl.1838.0019

Wheatstone, C. (1852). The Bakerian Lecture.—Contributions to the physiology of vision.—Part the second. On some remarkable, and hitherto unobserved, phenomena of binocular vision (continued). *Philosophical Transactions of the Royal Society of London*, *142*, 1–17. https://doi.org/10.1098/rstl.1852.0001

Yu, Y., Todd, J. T., & Petrov, A. A. (2021). Failures of stereoscopic shape constancy over changes of viewing distance and size for bilaterally symmetric polyhedra. *Journal of Vision*, *21*(6), 5. https://doi.org/10.1167/jov.21.6.5